\documentclass[%
 aip,
 amsmath,amssymb,
 reprint,%
]{revtex4-1}

\usepackage{graphicx}
\usepackage{dcolumn}
\usepackage{bm}
\usepackage{color,soul}

\usepackage[utf8]{inputenc}
\usepackage[T1]{fontenc}
\usepackage{mathptmx}
\usepackage{etoolbox}

\makeatletter
\def\@email#1#2{%
 \endgroup
 \patchcmd{\titleblock@produce}
  {\frontmatter@RRAPformat}
  {\frontmatter@RRAPformat{\produce@RRAP{*#1\href{mailto:#2}{#2}}}\frontmatter@RRAPformat}
  {}{}
}%
\makeatother
\begin{document}

\preprint{AIP/123-QED}

\title{Substrate conformal imprint fabrication process of synthetic antiferromagnetic nanoplatelets}
\author{J. Li} \email{j.li6@tue.nl}
\author{P. van Nieuwkerk}
\affiliation{Department of Applied Physics, Eindhoven University of Technology, P.O. Box 513, 5600 MB Eindhoven, The Netherlands}%
\author{M.A. Verschuuren}
\affiliation{SCIL nanoimprint solutions, Philips Research Laboratories, Eindhoven, The Netherlands}%
\author{B. Koopmans}
\author{R. Lavrijsen}
\affiliation{Department of Applied Physics, Eindhoven University of Technology, P.O. Box 513, 5600 MB Eindhoven, The Netherlands}%

\begin{abstract}
Methods to fabricate and characterize monodisperse magnetic nanoplatelets for fluid/bio-based applications based on spintronic thin-film principles are a challenge. This is due to the required top-down approach where the transfer of optimized blanket films to free particles in a fluid while preserving the magnetic properties is an uncharted field. Here, we explore the use of substrate conformal imprint lithography (SCIL) as a fast and cost-effective fabrication route. We analyze the size distribution of nominal 1.8 \textmu m and 120 nm diameter platelets and show the effect of the fabrication steps on the magnetic properties which we explain through changes in the dominant magnetization reversal mechanism as the size decreases. We show that SCIL allows for efficient large-scale platelet fabrication and discuss how application-specific requirements can be solved via process and material engineering. 
\end{abstract}

\maketitle
Magnetic particles have been widely used in bio-applications due to their ability to mechanically manipulate their surroundings remotely via externally applied magnetic fields. In particular, the utilization of magnetic torques induced via an externally rotating magnetic field is of interest for applications such as micro mixing \cite{gao2014chaotic}, cancer treatment \cite{chiriac2018fe,engelmann2018combining} and the manipulation of cells \cite{Dobson2008}. Superparamagnetic nanoparticles (SPNs) have traditionally been used in torque-related applications \cite{Erb2016}. However, due to their limited magnetic anisotropy and spherical shape, the translation of magnetic torque to mechanical torque is limited. Despite these limitations, particles with enhanced shape or magnetic anisotropy have been studied, e.g. NiFe nanodiscs with a vortex spin configuration \cite{Rozhkova2009,Kim2009}, and magnetic nanorods \cite{Martinez-Banderas2016}. Synthetic antiferromagnetic (SAF) nanoplatelets (NPs) with high perpendicular magnetic anisotropy (PMA) are among the most promising candidates \cite{Vemulkar2017c,Vemulkar2015a,Varvaro2019c,Mansell2017}. 

SAF NPs with PMA typically consist of a multilayer stack: Ta/Pt/Co/Pt/Ru/Pt/Co/Pt. The strong hybridization of the 3d-5d orbitals at Co and Pt interfaces induces a large PMA \cite{Carcia1998,Nakajima1998}. This large anisotropy and the fact that PMA induces a \textit{hard-plane} anisotropy (the plane of the disc) and an \textit{easy-axis} perpendicular to the the disc, are the key factors for effective magnetic-mechanical torque transduction \cite{Mansell2017}. The two ferromagnetic layers are antiferromagnetically coupled by the Ru layer through the Ruderman-Kittel-Kasuya-Yoshida (RKKY) interaction \cite{Ruderman1954,Kasuya1956,Yosida1957}. The SAF stack exhibits a zero net magnetic moment at zero applied magnetic field, preventing the aggregation of particles in liquid at zero field; a key requisite for applications. The Pt layers around the Ru layer tune the RKKY interaction and increase the PMA \cite{Lavrijsen2012b}. The high tunability of the PMA and RKKY and the freedom in shape and size of the platelets using top-down lithography methods make SAF NPs fascinating for remotely induced nanoscale torque applications. 

\begin{figure*}[ht]
\includegraphics[width=150mm,scale=0.5]{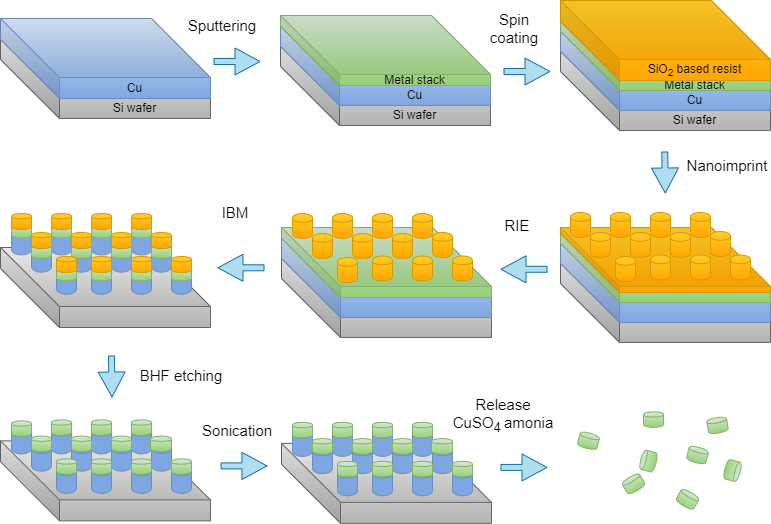}
\caption{\label{fig:fab}The schematic of the fabrication process of PMA-SAF NPs. (1) A sacrificial layer of Cu is sputtered on the Si wafer. (2) The SAF stack is sputtered on the Cu layer. (3) Spin-coat the resist. (4) Imprint. (5) Reactive ion etching (RIE) is used to etch the resist. (6) The metal layers are etched by ion beam milling (IBM). (7) The remaining resist is removed by buffered hydrogen fluoride (BHF) solution. (8) The re-deposition caused by the IBM process is removed by sonication.  (9) Finally, the NPs are released by dissolving the Cu layer in CuSO$_4$-ammonia solution}
\end{figure*}

However, one of the major issues is to fabricate monodisperse SAF NPs with high throughput and low cost. UV-lithography with a lift-off process has been reported to pattern 1.8 \textmu m diameter SAF NPs \cite{Vemulkar2015a}. However, due to the diffraction limit, it is hard to reach the sub-micron meter with conventional UV-lithography. Nanoimprint were used to fabricate SAF NPs with in-plane anisotropy at much smaller diameters down to 122 nm \cite{Hu2008a, Hu2011a}. Although smaller size can be achieved, the additive lift-off process has its native problem, i.e. it is difficult to obtain a uniform thickness when the critical dimensions reach the resist thickness. As the PMA-SAF system requires \AA ngstrom scale control of the layer thickness to stabilize the magnetic behavior, such additive methods cannot be used. Hence, a subtractive method is preferred where one can start with a blanket film on a wafer. Recently, nanosphere lithography; where polystyrene (PS) beads were used as hard masks, combined with an ion milling process was reported to produce NPs with different sizes \cite{Tiberto2015a,Goiriena-Goikoetxea2016,Welbourne2021b}. Nevertheless, this method suffers from non-uniform PS bead size and moreover, the yield depends on the distribution of the beads over a large area.

In this paper, we present a subtractive method based on substrate conformal imprint lithography (SCIL) to fabricate monodisperse SAF NPs. The stamp used in SCIL is composed of two rubber layers on a thin glass support (see Fig. S1). The SCIL technique is based on a difference between the in-plane stiffness of the glass which avoids pattern deformation over large areas, while the out-of-plane flexibility from the rubber layers allows conformal contact to underlying surface features \cite{Verschuuren2017}. With these properties, SCIL can be used to pattern large wafers up to 300 mm while keeping a uniform size of the features. In addition, this technique can be used to fabricate NPs from the nanometer to micrometer range and different shapes of the platelets can be thought off as the stamp used for the process can be custom made. Here we focus on 120 nm and 1.8 \textmu m diameter disc-shaped SAF NPs and a subtractive method using Ar ion beam milling (IBM). The magnetic properties of the discs after fabrication are studied and compared to literature, indicating that the SCIL fabrication route is a good candidate for large-scale PMA-SAF production.

The SCIL based fabrication process for the PMA-SAF platelets is outlined in Fig. \ref{fig:fab} (see supplementary material section 1 for details). We start with depositing a 2" Si wafer with a 30 nm sacriﬁcial Cu layer and the SAF stack using DC magnetron sputtering.  The basic SAF stack is [Ta(4)Pt(2)/CoB(0.8)/Pt(0.3)/Ru(0.8)/Pt(0.3)/CoB(0.8)/Pt(2)] with thickness in nanometers. For 1.8 \textmu m SAF NPs we use 5 repetitions of the basic stack and for 120 nm NPs we use one repetition. Then, we spin-coat the SCIL sol-gel resist and manually imprint the pillar structure using a custom SCIL imprint station, followed by sol-gel dependent hot plate bake and stamp removal. After transfer, the masks are etched by selective reactive ion etching (RIE) to open the area around the pillars. The metal stack is then etched by a non-selective Ar ion beam milling (IBM) step and followed by buffered HF (BHF) dip to remove the residual sol-gel resist on top of the nanoplatelets. During the IBM process, re-deposition on the masks can cause irregular side walls to grow around the NPs (see Fig. S2 (e-f)). To remove the re-deposited material, the sample is immersed in deionized (DI) water and sonicated for 20 minutes. Finally, the NPs are released in solution by dissolving the Cu layer in 1.5\%  CuSO$_4$- 10\% ammonia solution \cite{Hu2008a}.

\begin{figure}[ht]
\includegraphics[width=70mm,scale=0.25]{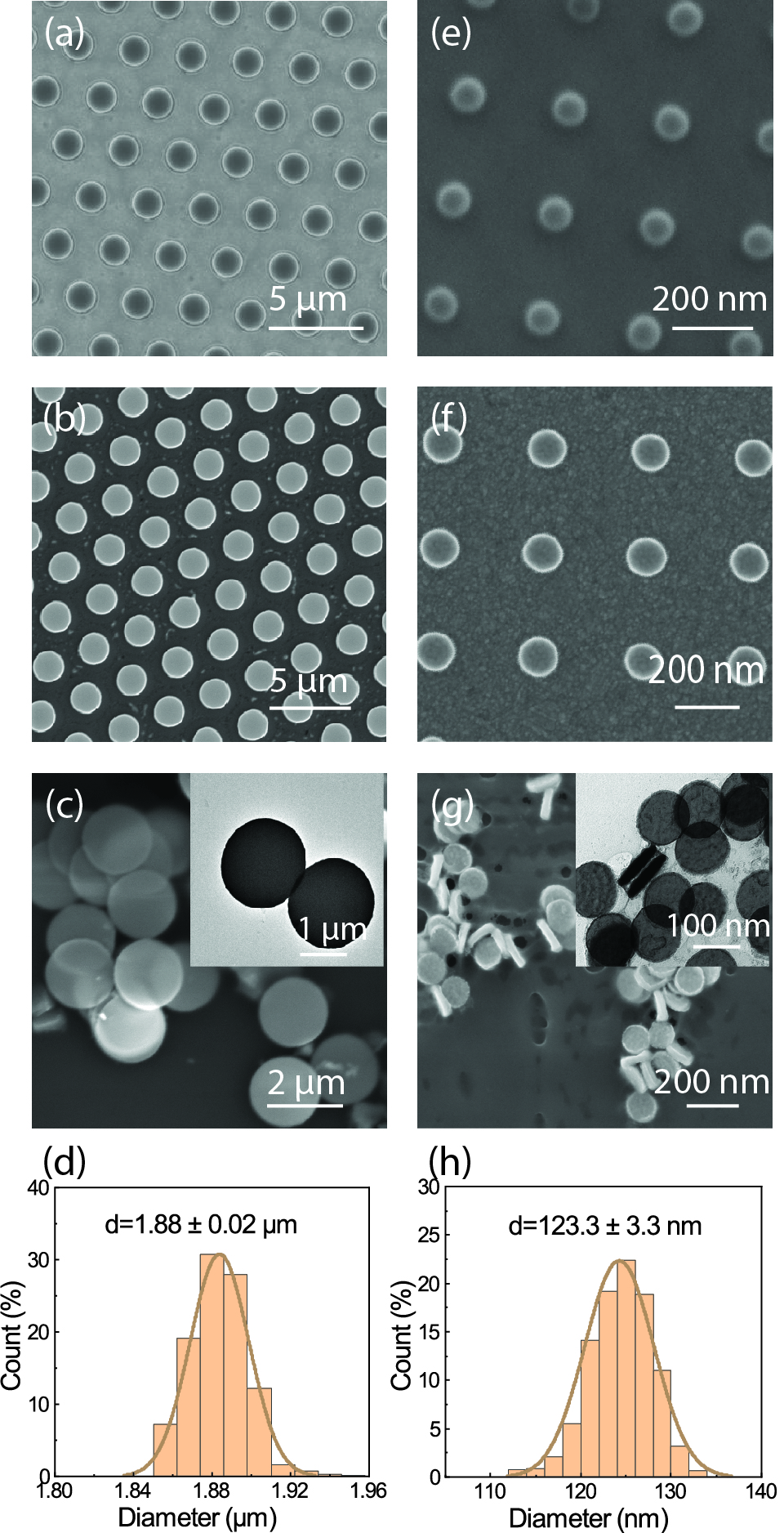}
\caption{\label{fig:sem}SEM images of 1.8 \textmu m diameter and 120 nm diameter SAF NPs (a, e)after imprinting, (b, f)before release, and (c, g)after release and dried on a Si substrate. The insert of (c) and (g) is the TEM image of 1.8 \textmu m and 120 nm diameter SAF NPs. (d) and (h) is the size distribution of released  1.8 \textmu m diameter and 120 nm diameter SAF nanoplatelets calculated from NPs before released.}
\end{figure}

The patterns of 1.8 \textmu m and 120 nm diameter NPs after nanoimprint, before release, and after release, were observed using scanning electron microscopy (SEM) as shown in Fig. \ref{fig:sem}. More details of the released sample preparation can be found in supplementary material section 2.A. After imprinting, monodisperse disc-shaped patterns are transferred successfully for both 1.8 \textmu m and 120 nm diameter NPs shown in Fig. \ref{fig:sem}(a) and \ref{fig:sem}(e). From Fig. \ref{fig:sem}(b-c) and Fig. \ref{fig:sem}(f-g), we see that the disc patterns are transferred into the metallic layer with uniform size after different etching processes and the NPs can be released without damaging the shape. The extracted size of the SAF NPs is shown in Fig. \ref{fig:sem}(d) and Fig. \ref{fig:sem}(h), which is 1.88 $\pm$ 0.02 \textmu m and 123.3 $\pm$ 3.3 nm, indicating a highly reproducible SCIL pattern transfer over the full 2" wafer area, with the size distribution approximately 1.1\% and 2.6\% (see supplementary material section 2.B for details related to size distribution calculation). 

Here, our used stamps have a packing density of 34 \%  and 11\% for 1.8  \textmu m and 120 nm NPs respectively  (see supplementary material section 1.F for details).  This packing density is not as high as the reported value of 50\% for 1 \textmu m discs and of 31\% for 100 nm discs fabricated through nanosphere lithography\cite{welbourne2021high}. However, the yield of SCIL which is determined by the specific stamp used for imprint can easily be increased by using a higher-packed mask and is not further addressed here.

\begin{figure*}[ht]
\includegraphics[width=150mm,scale=0.5]{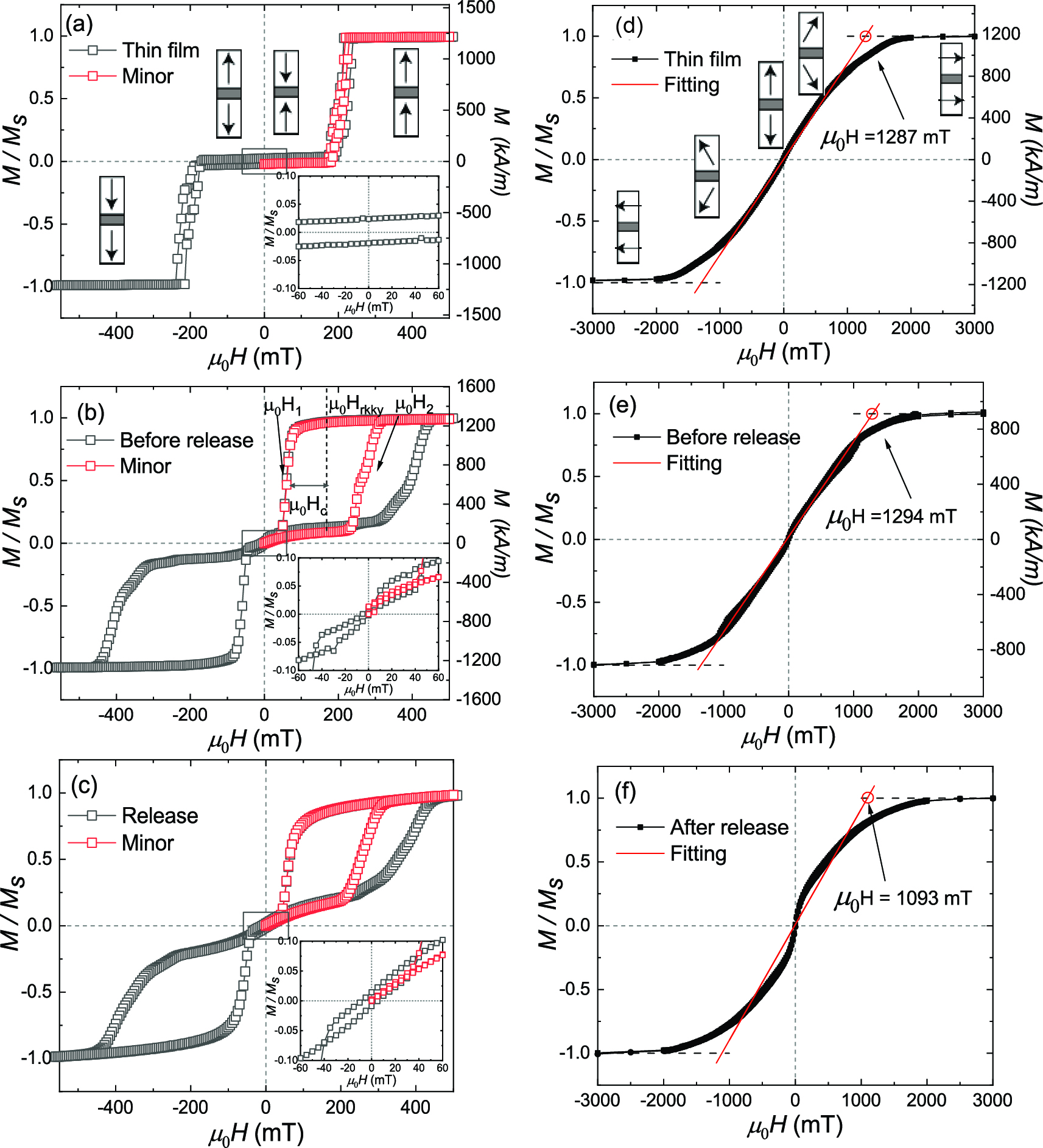}
\caption{\label{fig:squid} Hysteresis loops of the 1.8 \textmu m NPs measured through SQUID. The left column contains the hysteresis loops measured along the easy axis for (a) the as-deposited thin film (b) before release and (c) release in water and dried on a Si substrate. The inserts in (a-c) show the zoom-in region of the black box near the origin of the loop. The right column includes the hysteresis loops along the hard plane for (d) the as-deposited thin film (e) before release, and (f) after release. The arrows indicate the direction of the magnetization of the top and bottom ferromagnetic layers.}
\end{figure*}

\begin{figure*}[ht]
\includegraphics[width=150mm,scale=0.5]{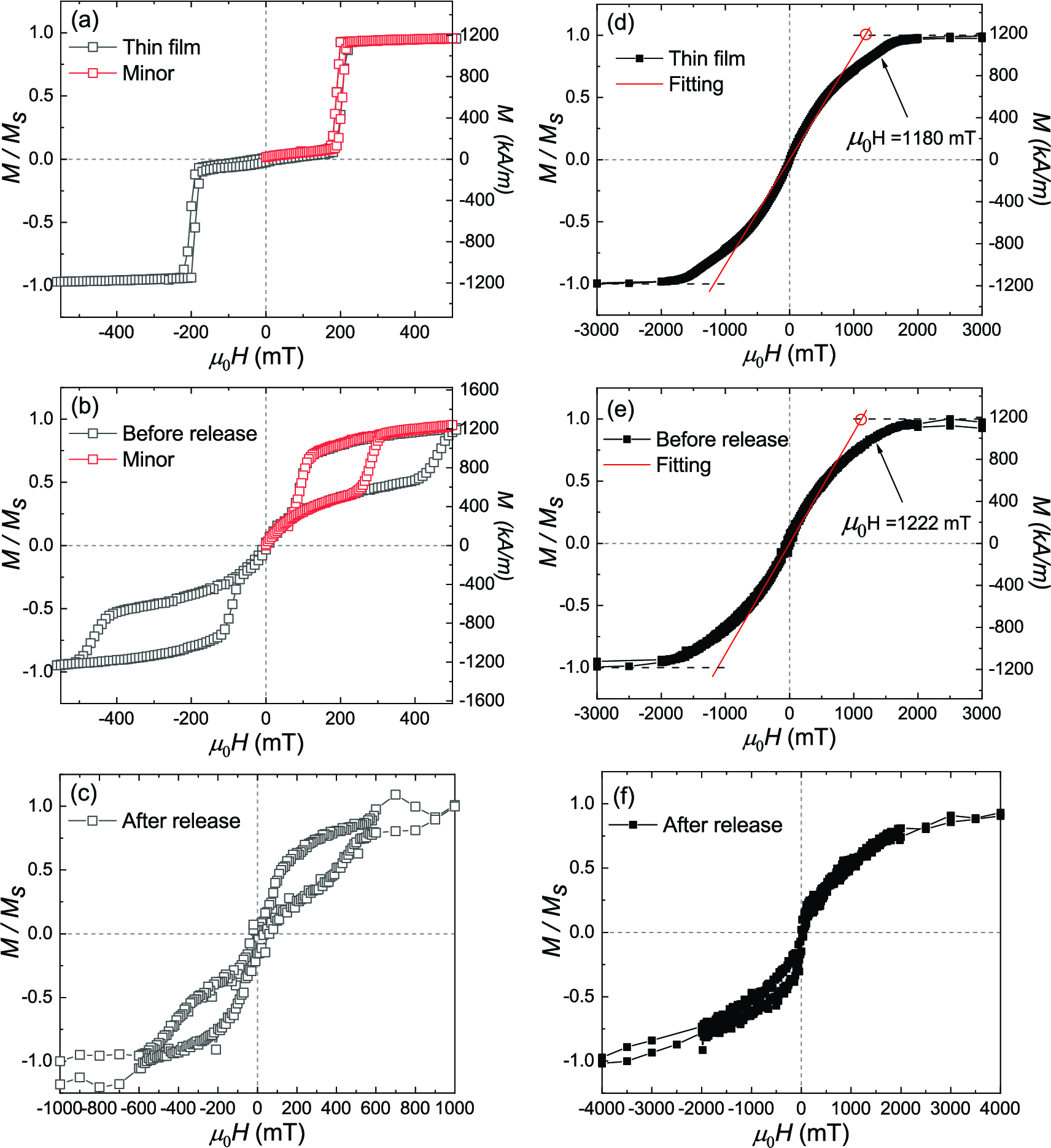}
\caption{\label{fig:squid120nm}   Hysteresis loops of the 120 nm NPs measured through SQUID. The left column contains the hysteresis loops measured along the easy axis for (a) the as-deposited thin film (b) before release and (c) after release. The right column includes the hysteresis loops along the hard plane for (d) the as-deposited thin film (e) before release, and (f) after release. }
\end{figure*}

To investigate the change in magnetic response due to the fabrication, the hysteresis loops of 1.8 \textmu m and 120 nm SAF NPs were measured by SQUID magnetometry as shown in Fig. \ref{fig:squid} and Fig. \ref{fig:squid120nm}. The left and right column of the figures show the hysteresis loops measured with the applied magnetic field perpendicular (easy-axis) and parallel (hard-plane) to the film plane respectively. To obtain the minor loop, the samples were first saturated in a positive field, then the magnetic field was decreased to zero and swept back to the positive saturation field. From the minor loop, the RKKY coupling field ($\mu_0H_{rkky}$) is defined as $\frac{\mu_0H_1 + \mu_0H_2}{2}$ and the coercivity ($\mu_0H_c$) is defined by $\frac{\mu_0H_2 -\mu_0H_1}{2}$ as shown in Fig \ref{fig:squid}(b) \cite{Lavrijsen2012b}.  

Let us first discuss the magnetic properties of 1.8 \textmu m NPs. The as-deposited blanket thin film is shown in Fig. \ref{fig:squid}(a). At a low magnetic field, the total magnetization is approximately zero, which is expected from the antiferromagnetic coupling of the top and bottom CoB layers in the basic stack and nearly equal magnetic moment of the two CoB layers. A small remnant moment at zero field can be observed in the inset Fig. \ref{fig:squid}(a), this is due to a slight thickness difference of the CoB layers during sequential growth \cite{Welbourne2021b}. Increasing the external field leads to a abrupt magnetization switch of the layer as expected from PMA-SAF samples in the spin-flip regime (i.e. where the PMA is much larger than the RKKY coupling) when the field is applied along the easy-axis \cite{Lavrijsen2012b}. This field we term the switching field, which depends on $\mu_0H_{rkky}$ and $\mu_0H_c$. From the minor loop we can extract $\mu_0H_{rkky}$ = 208 $\pm$ 10 mT and $\mu_0H_c$ = 6 $\pm$ 1 mT for the blanket film (see supplementary material section 3 for the details of error calculation). 

Let us now turn to the switching behavior of the patterned film of 1.8 \textmu m NPs before release as shown in Fig. \ref{fig:squid}(b). Ideally, the magnetic properties of the blanket film are propagated to the patterned NPs, however, two main differences can be observed; (1) $\mu_0H_c$ has increased from 6 mT to 102 mT, and the switches are smeared out in the field. (2) The coupling field $\mu_0H_{rkky}$ has reduced from 208 mT to 165 mT. The first observation can be explained by the dominant magnetic reversal mechanism (nucleation vs domain wall propagation \cite{Hu2005b,Hu2005c}) which for these PMA-SAF films typically depends on the number of defects per surface area. Reducing the area of the object (pattering), the chance of finding a defect per platelet reduces, hence this leads to the increase of the $\mu_0H_c$. Moreover, the sample area probed in the SQUID is around 4x4 mm$^2$ containing $ \sim 1.5\times$ 10$^6$ particles. The hysteresis observed is an ensemble response of all the NPs in the sample, from which the distribution of switching fields can be explained. The change in coupling field $\mu_0H_{rkky}$ has been observed before and is attributed to processing induced changes \cite{Welbourne2021}. Despite these differences, the typical SAF properties namely two distinctive switches and the well defined antiferromagnetic state at zero applied field are observed, which are similar to the blanket thin film. The saturation magnetization of the platelets remains constant at around 1300 kA/m between the blanket and patterned film (see Fig. \ref{fig:squid}(a) and \ref{fig:squid}(b)), which indicates that the fabrication process does not change the magnetic properties significantly. 

Let us now concentrate on the hysteresis loop measured on the released platelets as shown in Fig. \ref{fig:squid}(c) (see supplementary Section 2.A for sample preparation). As we cannot reliably quantify the number of platelets measured, we can only quantify the field response and not the saturation magnetization. Overall the observed response is similar to the NPs before release.  However, the hysteresis loop becomes more slanted and there is a slight increase in the switching distribution reflected in $\mu_0H_c$, which speculatively attribute to a distribution in the angle of alignment of the dried-in platelets relative to the applied field direction and possibly stray fields of piled up platelets (see also Fig. \ref{fig:sem}(c) where many platelets piled on top of each other). Overall, it is clear that the 1.8 \textmu m NPs before and after release keep their antiferromagnetic state at a low magnetic field and switch at high applied fields. The similar $\mu_0H_{rkky}$ and $\mu_0H_c$ indicate that the final release step does not degrade the SAF properties. On comparing the blanket film to the final released particles the most prominent change is found in the coercive field $\mu_0H_c$ which we attribute to the well-known size effect of patterning PMA films \cite{Thomson2006a}.

We will now move on to discuss the PMA of the NPs, which is the key factor of effective torque transduction \cite{Erb2016}. To demonstrate the PMA of the NPs, the hard-plane hysteresis loops are shown in the right column of Fig. \ref{fig:squid}. The deposited thin film, NPs before release and after release show nearly identical hysteresis loops. At zero field, the magnetization of the layers points antiparallel to each other leading to zero net magnetization. With increasing absolute field, the magnetization of the top and bottom CoB layers are tilted towards the applied field. Further increasing the field, the magnetizations tilt more and finally saturate. Here we define $\mu_0H_{sat}$ as the saturation field, which is the crossing point of the saturated state ($\frac{M}{M_s}=1$) and a linear fit of the data from -1000 mT to 1000 mT. To achieve the saturated state, the applied field should be large enough to overcome both the PMA and the RKKY interaction, from which $\mu_0H_{sat}$ can be defined as $\mu_0H_{sat}$ = $\mu_0H_{k}$ + 2$\mu_0H_{rkky}$, where $\mu_0H_k$ is the effective PMA field of the magnetic layer \cite{Mansell2017}. The perpendicular anisotropy energy ($K$) is given by $K = \frac{H_k M_s}{2} $, where $M_s$ is the saturation magnetisation of CoB. Both $\mu_0H_{rkky}$ and $M_s$ are obtained from the easy-axis hysteresis loops. From the equation above, the $K$ for the thin film sample, the NPs before release and after release are  $(5.4\pm0.4)\times10^5$, $(5.9\pm0.5)\times10^5$ and $(4.7\pm0.4)\times10^5$ $J/m^3$, respectively. We attribute the difference in $K$ of NPs before and after release to a spread in the alignment of the NPs relative to the applied magnetic field after drop-casting and drying. This directly affects the shape of the hysteresis loop where a reduced saturation field is to be expected. Overall the relatively small spread of $K$ $\approx 15\%$ denotes that the PMA is maintained during fabrication.

After discussing 1.8 \textmu m SAF NPs,  let's now examine the magnetic properties of 120 nm NPs depicted in Fig. \ref{fig:squid120nm}. First, we observe that all three samples exhibit SAF properties (see Fig. \ref{fig:squid120nm}(a-c)). The $\mu_0H_{rkky}$ is 197 $\pm$ 6 mT for blanket film and 189 $\pm$ 8 mT for NPs before release. After patterning, the $\mu_0H_{c}$ increased from 8 mT (blanket film) to 94 mT (before release). It is clear that the $\mu_0H_{rkky}$ and $\mu_0H_{c}$ of 120 nm NPs have similar values compared to the 1.8 \textmu m NPs, and their changes follow the same trend. The main difference is observed in the hysteresis loop of the released sample (see Fig. \ref{fig:squid120nm}(c)), where the switching fields are observed to be slanted and have a wider spread. This is due to the ill defined arrangement of dried NPs on the substrate (see Fig. \ref{fig:sem}(g)). This phenomenon is also reflected in the hard-plane hysteresis loop of released 120 NPs as shown in Fig. \ref{fig:squid120nm}(f), where the embedded switching behavior can be observed. Due to the misalignment, it is hard to quantitatively determine the $\mu_0H_{rkky}$, $\mu_0H_{c}$ and $K$ of released 120 nm NPs. Moving now on to consider the PMA of 120 nm NPs during fabrication, which we calculate from the hard-plane hysteresis loop shown in Fig. \ref{fig:squid120nm}(d-f). Here we do not include the $K$ of the released NPs. The $K$ values for the thin film sample and the NPs before release are $(4.8 \pm 0.4) \times10^5$ and $(5.2\pm 0.4)\times10^5$ $J/m^3$, respectively, which are comparable to the values of 1.8 \textmu m SAF NPs. Based on the hard-plane loop and the fact that the released step does not degradate the magnetic properties, although we cannot directly obtain the $K$ of released NPs, we can conclude that the PMA of 120 nm NPs remains at a high value before release and we expect that the high PMA remains even after the release step, which we will address in future work. In summary, after fabrication 120 NPs show SAF properties and the PMA does not decrease  .  

In conclusion, we show that SCIL can be used to fabricate PMA-SAF based disc-shaped platelets with a uniform diameter in the micrometer and sub-micrometer range. After fabrication, both 1.8 \textmu m and 120 nm NPs maintain their high PMA for high torque applications. A change in the magnetic response is observed which can be explained by the well-known change in the magnetic switching behavior when the area of PMA-SAF films changes. Our results pave the way for using SCIL imprint for the large-scale production of magnetic nanoplatelets using a subtractive method. 

\begin{acknowledgments}
The authors gratefully acknowledge Janne-Mieke Meijer and  Max Schelling from Eindhoven University of Technology for their assistance with the TEM measurements.
\end{acknowledgments}

\section*{AUTHOR DECLARATIONS}
\subsection*{Conflict of Interest}
The authors have no conflicts to disclose.
\subsection*{DATA AVAILABILITY}
The data that support the findings of this study are available.



\bibliography{aipsamp}
\end{document}